\documentclass[prl,twocolumn,superscriptaddress,showpacs,amsmath,amssymb]{revtex4}
\usepackage{graphicx}
\usepackage{bm}
\usepackage{amsmath,amssymb}
\begin{document}
\title{Not all physical errors can be linear CPTP maps
in a correlation space
}  
\author{Tomoyuki Morimae}
\affiliation{
Universit\'e Paris-Est Marne-la-Vall\'ee, 77454 Marne-la-Vall\'ee
Cedex 2, France}
\author{Keisuke Fujii}
\affiliation{Graduate School of Engineering Science,
Osaka University, Toyonaka, Osaka 560-8531, Japan}
\date{\today}
            
\begin{abstract}
In the framework of quantum computational tensor network, which is
a general framework of measurement-based quantum computation, 
the resource many-body state
is represented in a tensor-network form, and universal quantum computation
is performed in a virtual linear space, which is called a
correlation space, where tensors live.
Since any unitary operation, state preparation, and the projection
measurement in the computational basis can be simulated
in a correlation
space, it is natural to expect that fault-tolerant
quantum circuits
can also be simulated in a correlation space.
However, we point out that 
not all physical errors  
on physical qudits 
appear as linear completely-positive trace-preserving errors 
in a correlation space.
Since
the theories of fault-tolerant quantum circuits known so far
assume such noises,
this means that
the simulation of fault-tolerant quantum circuits in
a correlation space is not so straightforward for
general resource states.

\end{abstract}
\pacs{03. 67. -a}
\maketitle  

{\it Introduction}.---
Quantum many-body states, which have long been central research
objects in condensed matter physics, statistical physics,
and quantum chemistry,
are now attracting the renewed interest in quantum information science
as fundamental resources for quantum information processing.
One of the most celebrated examples
is one-way 
quantum computation~\cite{one-way,one-way2,one-way3}. 
Once the highly-entangled
many-body state which is called the cluster state
is prepared, universal quantum computation is possible with
adaptive local measurements on each qubit.
Recently, the concept of 
quantum computational tensor network (QCTN)~\cite{Gross1,Gross2,Gross3},
which is the general framework of measurement-based
quantum computation on quantum many-body states,
was proposed. 
This novel framework has enabled us to understand how 
general measurement-based quantum computation is 
performed on many other resource states
beyond the cluster state.
The most innovative feature of QCTN is that the resource
state is represented in   
a tensor network form~\cite{Fannes,Verstraete,Cirac}, 
and universal quantum computation
is performed in the virtual linear space where tensors live.
For example, let us consider the one-dimensional open-boundary chain of $N$ qudits 
in the matrix product form
\begin{eqnarray}
|\Psi\rangle\equiv
\sum_{k_1=0}^{d-1}...\sum_{k_N=0}^{d-1}
\langle L|A[k_N]...A[k_1]|R\rangle
|k_N,...,k_1\rangle,
\label{MPS}
\end{eqnarray}
where 
$\{|0\rangle,...,|d-1\rangle\}$ is a certain basis in 
the $d$-dimensional
Hilbert space ($2\le d<\infty$),
$|L\rangle$ and $|R\rangle$ are $D$-dimensional complex
vectors, and $\{A[0],...,A[d-1]\}$ are $D\times D$ complex matrices.
Let us also define the projection measurement $\mathcal{M}_{\theta,\phi}$ 
on a single physical qudit by
${\mathcal M}_{\theta,\phi}\equiv\big\{|\alpha_{\theta,\phi}\rangle,
|\beta_{\theta,\phi}\rangle,
|2\rangle,...,|d-1\rangle \big\}$,
where
$|\alpha_{\theta,\phi}\rangle
\equiv\cos\frac{\theta}{2}|0\rangle+e^{i\phi}\sin\frac{\theta}{2}|1\rangle$,
$|\beta_{\theta,\phi}\rangle
\equiv\sin\frac{\theta}{2}|0\rangle-e^{i\phi}\cos\frac{\theta}{2}|1\rangle$,
$0<\theta<\pi$, and $0\le \phi < 2\pi$.
If we do the measurement ${\mathcal M}_{\theta,\phi}$ on the first 
physical qudit 
of Eq.~(\ref{MPS}) and if
the first physical qudit is projected onto, for example, 
$|\alpha_{\theta,\phi}\rangle$
as a result of this measurement,
the state Eq.~(\ref{MPS}) becomes
\begin{eqnarray*}
\sum_{k_2,...,k_N}
\langle L|A[k_N]...A[k_2]\frac{A[\alpha_{\theta,\phi}]}
{\|A[\alpha_{\theta,\phi}]\|}
|R\rangle
|k_N,...,k_2\rangle\otimes|\alpha_{\theta,\phi}\rangle,
\end{eqnarray*}
where 
$A[\alpha_{\theta,\phi}]\equiv\cos\frac{\theta}{2}A[0]+
e^{-i\phi}\sin\frac{\theta}{2}A[1]$.
Then, we say ``the operation 
$|R\rangle\to
\frac{A[\alpha_{\theta,\phi}]}{\|A[\alpha_{\theta,\phi}]\|}
|R\rangle$
is implemented in the correlation space".
In particular,
if $A[0]$, $A[1]$, $\theta$, and $\phi$
are appropriately chosen in such a way that 
$A[\alpha_{\theta,\phi}]$ is proportional to a unitary,
we can ``simulate" the unitary evolution 
$\frac{A[\alpha_{\theta,\phi}]}{\|A[\alpha_{\theta,\phi}]\|}|R\rangle$
of the vector $|R\rangle$ in
the virtual linear space where $A$'s, $|R\rangle$, and $|L\rangle$ live.
This virtual
linear space is called the correlation space~\cite{Gross1,Gross2,Gross3}.
The core of 
QCTN 
is this ``virtual quantum computation" in the correlation space. 
If the correlation space has a sufficient structure
and if $A$'s, $|L\rangle$, and $|R\rangle$ are appropriately chosen,
we can ``simulate" universal quantum circuit in the correlation 
space~\cite{Gross1,Gross2,Gross3,Download,FM}.

For the realization of a scalable quantum computer, 
a theory of fault-tolerant (FT)
quantum computation~\cite{Shor,Aharonov,Kitaev,Knill,Terhal} is necessary.
In fact, several researches have been performed
on FT quantum computation
in the one-way model~\cite{one-way3,ND05,AL06,Silva,Fujii1,Fujii2}.
However, 
there has been no result about a theory of FT quantum computation 
on general QCTN~\cite{recently}.
In particular, there is severe lack of knowledge about
FT quantum computation
on resource states with $d\ge3$.
It is necessary to  
consider 
resource states with $d\ge3$
if we want to enjoy the cooling preparation
of a resource state
and the energy-gap protection of measurement-based quantum computation
with a physically natural Hamiltonian, 
since no genuinely entangled qubit state can be
the unique ground state of a two-body frustration-free
Hamiltonian~\cite{no-go}. 

One straightforward way of implementing FT quantum computation
on QCTN
is to encode physical qudits with a quantum error correcting
code:
$\sum_{k_1=0}^{d-1}...\sum_{k_N=0}^{d-1}
\langle L|A[k_N]...A[k_1]|R\rangle
|\tilde{k}_N,...,\tilde{k}_1\rangle$,
where $|\tilde{k}_i\rangle$ ($i=1,...,N$) is the encoded version
of $|k_i\rangle$ (such as $|\tilde{0}\rangle=|000\rangle$ and $|\tilde{1}\rangle=|111\rangle$, etc.)
In fact, this strategy was taken in Refs.~\cite{Fujii1,Fujii2}
for the one-way model $(d=2)$,  
and it was shown there that a FT construction of 
the encoded cluster state
is possible.
For $d\ge3$, however,
such a strategy is difficult, since
theories of  
quantum error correcting codes 
and 
FT preparations of 
the encoded resource 
state  
are less developed for $d\ge3$.
Furthermore, if we encode physical qudits with a quantum error
correcting code, the parent
Hamiltonian should no longer be two-body interacting one.

The other way of implementing FT quantum computation
on QCTN
is to simulate FT quantum circuits
in the correlation space.
Since any unitary operation, state preparation,
and the projective measurement 
in the computational basis
can be simulated in a correlation space (for a precise discussion
about the possibility of the measurement, see Ref.~\cite{FM}),  
it is natural to expect that FT quantum circuits
can also be simulated in a correlation space.
An advantage of this strategy is that 
theories of FT quantum circuits
for qubit systems are well developed~\cite{Shor,Aharonov,Kitaev,Knill,Terhal}. 
In fact, this strategy was taken in Refs.~\cite{one-way3,
ND05,AL06} for the one-way model $(d=2)$.
They simulated FT quantum circuits
on the one-way model.

In this paper, however,
we point out that it is not so straightforward to simulate
FT quantum circuits  
in a correlation space in general.
We first consider the simulation of quantum circuits in the correlation
space of pure matrix product states (post-measurement conditional states).
We show that if 
$d\ge3$
not all physical errors on physical qudits
appear as linear completely-positive trace-preserving (CPTP) 
errors in the correlation space.
Since all theories of FT quantum circuits
known so far
assume such noises~\cite{Shor,Aharonov,Kitaev,Knill,Terhal},
this means that it is not so straightforward to apply these FT theories
to quantum circuits simulated in the correlation space
of pure matrix product states.

We therefore next consider another way
of simulating quantum
circuits in the correlation space by mixing measurement results.
For the cluster state and the tricluster state~\cite{tricluster},
such a mixing strategy well works:
all CPTP errors on a physical qubit (or qudit) can be CPTP errors
in the correlation space.
However, this is not the case for other general resource states.
As an example, we consider the one-dimensional AKLT state~\cite{Brennen,AKLT}, and see that
not all physical errors on a physical qutrit
can be linear CPTP errors in the correlation space even if we mix measurement results.
This suggests that even if we mix measurement results,
like the cluster model,
the simulation of FT quantum circuits in the correlation space
of a general resource state
is not so straightforward.

{\it Simulation with pure states}.---
First, let us we consider the simulation of quantum circuits in the correlation
space of pure matrix product states.
We show 
that if $d\ge3$
not all physical errors on physical qudits
appear as linear CPTP 
errors in the correlation space.

Since the MPS, Eq.~(\ref{MPS}), is a resource state for measurement-based
quantum computation,
we can assume without loss of generality that 
$A[\alpha_{\theta,\phi}]$, $A[\beta_{\theta,\phi}]$, $A[2]$, 
$A[3]$, ..., $A[d-1]$
are unitary
up to constants:
$A[\alpha_{\theta,\phi}]=c_\alpha U_\alpha$,
$A[\beta_{\theta,\phi}]=c_\beta U_\beta$,
$A[2]=c_2 U_2$,
$A[3]=c_3 U_3$,
...
$A[d-1]=c_{d-1} U_{d-1}$,
where $c_\alpha$, $c_\beta$, $c_2$, ... $c_{d-1}$ are real positive numbers,
$U_\alpha$, $U_\beta$, $U_2$, ..., $U_{d-1}$ are unitary operators,
and
$A[\beta_{\theta,\phi}]\equiv\sin\frac{\theta}{2}A[0]-e^{-i\phi}\cos\frac{\theta}{2}A[1]$.
This means that
any operation implemented in the correlation space
by the measurement ${\mathcal M}_{\theta,\phi}$ on a single
physical qudit of Eq.~(\ref{MPS}) is unitary.
Note that this assumption is reasonable, since otherwise 
Eq.~(\ref{MPS}) does not seem to be useful as a resource for measurement-based quantum computation.
In fact, all known resource states so far~\cite{one-way,one-way2,one-way3,Gross1,Gross2,Gross3,Brennen,Caimagnet,tricluster,Miyake2dAKLT},
including the cluster state and the AKLT state, 
satisfy this assumption by appropriately rotating each local physical basis.
Furthermore, we can take $c_\alpha$, $c_\beta$, $c_2$,..., $c_{d-1}$ such that 
$C\equiv c_\alpha^2+c_\beta^2+\sum_{k=2}^{d-1}c_k^2=1$,
since
$\sum_{k_1,...,k_N}\langle L|A[k_N]...A[k_1]|R\rangle
|k_N,...,k_1\rangle
=\sqrt{C}^N
\sum_{k_1,...,k_N}\langle L|\frac{A[k_N]}{\sqrt{C}}...\frac{A[k_1]}{\sqrt{C}}|R\rangle
|k_N,...,k_1\rangle
$
and we can redefine $A[k_i]/\sqrt{C}\to A[k_i]$.

{\bf Theorem}:
If $d\ge3$, there exists a single-qudit CPTP error $\mathcal E$ 
which has the following property:
assume that $\mathcal E$ is applied on a single physical qudit of 
Eq.~(\ref{MPS}).
If the measurement ${\mathcal M}_{\theta,\phi}$ is performed on 
that affected qudit, 
a non-TP operation is implemented in the correlation space.

{\bf Proof}:
In order to show Theorem, let us assume that  
\begin{eqnarray}
\mbox{There is no such $\mathcal E$}. 
\label{assumption}
\end{eqnarray}
We will see that this assumption leads to
the contradiction that $d\le2$.

First, let us consider 
the state
\begin{eqnarray}
(I^{\otimes N-1}\otimes U_{1\leftrightarrow 2})
|\Psi\rangle,
\label{error1}
\end{eqnarray}
where  
$U_{a\leftrightarrow b}\equiv
|a\rangle\langle b|+|b\rangle\langle a|
+I
-|a\rangle\langle a|-|b\rangle\langle b|$
is the unitary error which exchanges $|a\rangle$ and $|b\rangle$,
and $I$ is the identity operator on a single qudit.
In Eq.~(\ref{error1}), 
the error $U_{1\leftrightarrow2}$ is applied on the first physical
qudit of $|\Psi\rangle$.
If we do the measurement ${\mathcal M}_{\theta,\phi}$ on the first physical
qudit of
Eq.~(\ref{error1}),
and if the measurement result is $|2\rangle$,
Eq.~(\ref{error1})
becomes
\begin{eqnarray}
\sum_{k_2,...,k_N}
\langle L|A[k_N]...A[k_2]\frac{A[1]}{\|A[1]\|}|R\rangle
|k_N,...,k_2\rangle\otimes|2\rangle.
\label{error1_after}
\end{eqnarray}
In other words, the operation 
$|R\rangle\to
\frac{A[1]}{\|A[1]\|}|R\rangle$
is implemented in the correlation space.
By the assumption Eq.~(\ref{assumption}),
this operation should work as a TP operation in the correlation space.
Therefore,
\begin{eqnarray}
\frac{A^\dagger[1]}{\|A[1]\|}
\frac{A[1]}{\|A[1]\|}=I.
\label{error1_result}
\end{eqnarray}
By taking $\eta\equiv\|A[1]\|^2$,
we obtain
\begin{eqnarray}
A^\dagger[1]A[1]=\eta I.
\label{error1_result}
\end{eqnarray}

Second, let us consider the measurement ${\mathcal M}_{\theta,\phi}$ on 
the first physical qudit of
$(I^{\otimes N-1}\otimes U_{0\leftrightarrow 2}V^s)
|\Psi\rangle$,
where  
$s\in\{0,1,...,d-1\}$, 
$V\equiv\sum_{p=0}^{d-1}e^{-i\omega p}|p\rangle\langle p|$
is a unitary phase error,
and
$\omega\equiv 2\pi/d$.
If the measurement result is $|\alpha_{\theta,\phi}\rangle$,
$(e^{-2is\omega}\cos\frac{\theta}{2}A[2]
+e^{-i(\phi+s\omega)}\sin\frac{\theta}{2}A[1]
)/\sqrt{\gamma}$
is implemented in the correlation space,
where
$\sqrt{\gamma}\equiv
\|e^{-2is\omega}\cos\frac{\theta}{2}A[2]
+e^{-i(\phi+s\omega)}\sin\frac{\theta}{2}A[1]
\|$.
By the assumption Eq.~(\ref{assumption}), this should
work as a TP operation in the correlation space.
Therefore,
$\gamma I=\cos^2\frac{\theta}{2}A^\dagger[2]A[2]+
\sin^2\frac{\theta}{2}A^\dagger[1]A[1]
+\frac{1}{2}
\sin\theta
(e^{-i(\phi-s\omega)}A^\dagger[2]A[1]
+e^{i(\phi-s\omega)}A^\dagger[1]A[2])$.
By the assumption that all $A$'s are proportional to unitaries, 
$A^\dagger[2]A[2]=\xi I$,
where $\xi\equiv\|A[2]\|^2$.
Furthermore, as we have shown, $A^\dagger[1]A[1]=\eta I$ (Eq.~(\ref{error1_result})).
Therefore,
\begin{eqnarray}
\gamma' I=
e^{-i(\phi-s\omega)}A^\dagger[2]A[1]+
e^{i(\phi-s\omega)}A^\dagger[1]A[2],
\label{e3}
\end{eqnarray}
where 
$\gamma'\equiv \frac{2}{\sin\theta}(\gamma-\xi\cos^2\frac{\theta}{2}
-\eta\sin^2\frac{\theta}{2})$.

Finally, let us consider 
the measurement ${\mathcal M}_{\theta,\phi}$ on 
the first physical qudit of
$(I^{\otimes N-1}\otimes U_{0\leftrightarrow 1}U_{0\leftrightarrow 2}V^t)
|\Psi\rangle$,
where $t\in\{0,1,...,d-1\}$.
If the measurement result
is $|\alpha_{\theta,\phi}\rangle$,
$(e^{-it\omega}\cos\frac{\theta}{2} A[1]
+e^{-i\phi-2it\omega}\sin\frac{\theta}{2} A[2])/\sqrt{\delta}$
is implemented in the correlation space,
where
$\sqrt{\delta}\equiv\|e^{-it\omega}\cos\frac{\theta}{2} A[1]
+e^{-i\phi-2it\omega}\sin\frac{\theta}{2} A[2]\|$.
By the assumption Eq.~(\ref{assumption}), 
this should also work as a TP operation in the correlation
space. Therefore,
\begin{eqnarray}
\delta' I=
e^{i(\phi+t\omega)}A^\dagger[2]A[1]
+e^{-i(\phi+t\omega)}A^\dagger[1]A[2],
\label{e4}
\end{eqnarray}
where 
$\delta'\equiv\frac{2}{\sin\theta}(\delta-\xi\sin^2\frac{\theta}{2}
-\eta\cos^2\frac{\theta}{2})$.

From Eqs.~(\ref{e3}) and (\ref{e4}), 
$\epsilon I=
[
e^{-2i(\phi-s\omega)}-e^{2i(\phi+t\omega)}
]A^\dagger[2]A[1]$,
where 
$\epsilon\equiv e^{-i(\phi-s\omega)}\gamma'-e^{i(\phi+t\omega)}\delta'$.

Let us assume that
$e^{-2i(\phi-s\omega)}-e^{2i(\phi+t\omega)}\neq0$.
Then,
$\epsilon' I=A^\dagger[2]A[1]$,
where 
$\epsilon'\equiv \epsilon/(
e^{-2i(\phi-s\omega)}-e^{2i(\phi+t\omega)})$.
If $\epsilon'=0$,
$A^\dagger[2]A[1]=0$, which means $A[1]=0$ since $A[2]$ is unitary
up to a constant.
Therefore, $\epsilon'\neq0$.
In this case,
$A[1]=\epsilon'' A[2]$ for certain $\epsilon''\neq0$, 
since $A[2]$ is unitary up to a constant~\cite{exclude}.
Hence 
$e^{-2i(\phi-s\omega)}-e^{2i(\phi+t\omega)}=0$.
This means
\begin{eqnarray}
2\phi+(t-s)\omega=r_{s,t}\pi,
\label{theorem}
\end{eqnarray}
where $r_{s,t}\in\{0,1,2,3,...\}$.
Let us take $t=s=0$.
Then, Eq.~(\ref{theorem}) gives
$\phi=r_{0,0}\frac{\pi}{2}~~~(r_{0,0}\in\{0,1,2,...\})$.
Let us take $s=1$, $t=0$.
Then, Eq.~(\ref{theorem}) gives
$\phi=\frac{\pi}{d}+r_{1,0}\frac{\pi}{2}~~~(r_{1,0}\in\{0,1,2,...\})$.
In order to satisfy these two equations at the same time,
there must exist $r_{0,0}$ and $r_{1,0}$
such that
$r_{0,0}\frac{\pi}{2}
=\frac{\pi}{d}+r_{1,0}\frac{\pi}{2}$.
If $r_{0,0}=r_{1,0}$, then $0=1/d$ which means $d=\infty$.
Therefore
$r_{0,0}\neq r_{1,0}$.
Then we have
$d=2/(r_{0,0}-r_{1,0})
\le 2$,
which is the contradiction.~$\blacksquare$

One might think that if 
we rewrite the post-measurement state
Eq.~(\ref{error1_after})
as
\begin{eqnarray*}
\sum_{k_2,..,k_N}
\langle L|A[k_N]...A[k_2]
\frac{A[1]}
{\|A[1]|R\rangle\|}
|R\rangle
|k_N,...,K_2\rangle\otimes|2\rangle
\end{eqnarray*}
and redefine the operation implemented in the correlation space
as
$|R\rangle\to
\frac{A[1]}
{\|A[1]|R\rangle\|}
|R\rangle$,
the TP-ness is recovered in the correlation space.
However, in this case, the non-lineally appears
unless 
$A^\dagger[1]A[1]\propto I$,
and therefore if we require the linearity in the correlation space,
we obtain the same contradiction.

In short,
if $d\ge3$
not all physical errors on physical qudits
appear as linear CPTP 
errors in the correlation space of pure matrix product 
states~\cite{intuition}.

{\it Simulation by mixing measurement results}.---
We have seen that
if we simulate quantum circuits in the correlation space
of pure matrix product states,
not all physical errors on a physical qudit can be linear 
CPTP errors in the correlation space.
Therefore we must simulate quantum circuits in the correlation space
with another method:
we consider the simulation by mixing
measurement results.

Before studying a concrete example,
let us consider the effect of a CPTP error on general resource states.
Let us assume that a CPTP error 
$\rho\to\sum_{j=1}^wF_j\rho F_j^\dagger$,
where $\sum_{j=1}^wF_j^\dagger F_j=I$,
occurs on the first physical
qudit of $|\Psi\rangle$.
If we measure the first physical qudit in a certain basis $\{|m_s\rangle\}$,
we obtain~\cite{details}
$\sum_s
\sum_j
W(
E_{j,s}|R\rangle)_2
\otimes
|m_s\rangle\langle m_s|$,
where
$E_{j,s}\equiv
\sum_{k}A[k]
\langle m_s|F_j|k\rangle$
and
\begin{eqnarray*}
&&
W(|\psi\rangle)_r\equiv
\sum_{k_r,...,k_N}
\sum_{k'_r,...,k'_N}
\langle L|A[k_N]...A[k_r]|\psi\rangle\\
&&\langle\psi|A^\dagger[k'_r]...A^\dagger[k'_N]|L\rangle
|k_r,...,k_N\rangle\langle k'_r,...,k'_N|.
\end{eqnarray*}
If we trace out $|m_s\rangle$,
we obtain
$\sum_s\sum_jW(E_{j,s}|R\rangle)_2$.
This means that the map
\begin{eqnarray}
|R\rangle\langle R|\to
\sum_{s,j}E_{j,s}|R\rangle\langle R|E^\dagger_{j,s}
\label{map}
\end{eqnarray}
is implemented in the correlation space.
Since we can show~\cite{details}
$\sum_{j,s}E^\dagger_{j,s}
E_{j,s}=I$,
the map Eq.~(\ref{map}) is CPTP.

In general, we must do feed-forwarding before
tracing out $|m_s\rangle$ in order
to deterministically implement quantum gates in the correlation space.
As is shown in Refs.~\cite{AL06,details},
all CPTP errors on a physical qubit (or qudit) can be CPTP errors 
in the correlation space of the cluster state
and the tricluster state even if we do the feed-forwarding
in this mixing strategy.
However, it is not the case for other general resource states~\cite{GrossH}.
As an example, let us consider the one-dimensional AKLT state,
where $d=3$, 
$A[0]=\frac{1}{\sqrt{3}}X$,
$A[1]=\frac{1}{\sqrt{3}}XZ$,
$A[2]=\frac{1}{\sqrt{3}}Z$.

If a CPTP error occurs on the first physical qutrit,
and if we do the usual measurement-based quantum computation
on the AKLT state,
we obtain~\cite{details}
\begin{eqnarray}
&&
\sum_j
\sum_{p=0}^1
\sum_{q=0}^1
\Big(\nonumber\\
&&\sum_{(s_1,...,s_r)\in S_{p,q}^r}
W(Q_r(s_1,...,s_r)...Q_2(s_1,s_2)E_{j,s_1}|R\rangle)_{r+1}\nonumber\\
&&+
h(p,q,r)W(Z^{r-1}E_{j,2}|R\rangle)_{r+1}
\Big)
\otimes
\eta(p,q),
\label{kekka}
\end{eqnarray}
where $r$ is the number of measurements (since the 
gate operation is non-deterministic in AKLT model, we must
repeat measurements until we near-deterministically implement
the desired gate operation),
and
$S_{p,q}^r$ is the set of
measurement outcomes
$(s_1,...,s_r)\in\{0,1,2\}^r\setminus(2,...,2)$
such that
$f(s_1,...,s_r)=p$ and $g(s_1,...,s_r)=q$.
Here, $f(s_1,...,s_r)=\oplus_{i=1}^r(\delta_{s_i,0}\oplus\delta_{s_i,1})$
and $g(s_1,...,s_r)=\oplus_{i=1}^r(\delta_{s_i,1}\oplus\delta_{s_i,2})$.
Also, 
$Q_k(s_1,...,s_{k-1},0)=Xe^{iZ\theta/2}$,
$Q_k(s_1,...,s_{k-1},1)=XZe^{iZ\theta/2}$,
$Q_k(s_1,...,s_{k-1},2)=Z$
for $s_1=...=s_{k-1}=2$,
and
$Q_k(s_1,...,s_{k-1},0)=X$,
$Q_k(s_1,...,s_{k-1},1)=XZ$,
$Q_k(s_1,...,s_{k-1},2)=Z$
for other $s_i$'s.
We also define
$h(p,q,r)=\delta_{p,0}\delta_{q,0}$
if $r$ is even, and
$h(p,q,r)=\delta_{p,0}\delta_{q,1}$
if $r$ is odd.
Finally, $\eta(0,0)$, $\eta(0,1)$, $\eta(1,0)$, and $\eta(1,1)$
are mutually orthogonal states,
which record Pauli byproducts.
The first term of Eq.~(\ref{kekka}) corresponds to the mixture
of successful measurement results (desired rotation is implemented)
and the second term corresponds to the failed measurement
results (the desired rotation is not implemented.)

Equation~(\ref{kekka}) means that
for fixed $p$ and $q$, the map
\begin{eqnarray}
&&
|R\rangle\langle R|\to\nonumber\\
&&\sum_{(s_1,...,s_r)\in S_{p,q}^r}
\sum_j
\tilde{Q}(s_1,...,s_r,j)|R\rangle\langle R|
\tilde{Q}^\dagger(s_1,...,s_r,j)\nonumber\\
&&+
h(p,q,r)\sum_j
Z^{r-1}E_{j,2}|R\rangle\langle R|E^\dagger_{j,2}Z^{r-1}
\label{AKLTmap}
\end{eqnarray}
is implemented in the correlation space,
where
$\tilde{Q}(s_1,...,s_r,j)\equiv
Q_r(s_1,...,s_r)...Q_2(s_1,s_2)E_{j,s_1}$.

For example, 
let us consider the error with $w=1$
and
\begin{eqnarray*}
F_1=
U_{{\mathcal M}_{\theta,\phi}}
\Big(
\frac{|0\rangle+|1\rangle}{\sqrt{2}}\langle0|
-\frac{|0\rangle-|1\rangle}{\sqrt{2}}\langle1|
+|2\rangle\langle2|\Big),
\end{eqnarray*}
where $U_{{\mathcal M}_{\theta,\phi}}
\equiv|\alpha_{\theta,\phi}\rangle\langle0|
+|\beta_{\theta,\phi}\rangle\langle1|+\sum_{k=2}^{d-1}|k\rangle\langle k|$.
Then,
if $p=1$ and $q=0$,
we can show that~\cite{details}
\begin{eqnarray*}
\sum_{(s_1,...,s_r)\in S_{p,q}^r}
\sum_j
\tilde{Q}^\dagger(s_1,...,s_r,j)
\tilde{Q}(s_1,...,s_r,j)
\end{eqnarray*}
is $\alpha I+\frac{2}{3}|1\rangle\langle1|$
if $r$ is odd,
and $\beta I+\frac{2}{3}|0\rangle\langle0|$
if $r$ is even,
where $\alpha$ and $\beta$ are certain positive numbers~\cite{details}.
This means that the map Eq.~(\ref{AKLTmap})
is not linear CPTP.

{\it Conclusion}.---
In this paper, we have studied how physical errors on a physical qudit
appear in the correlation space.
We have shown that if $d\ge3$ not all physical errors
can be linear CPTP errors in the correlation space of 
pure matrix product states.
We have also shown that even if we mix the measurement
results, not all physical errors are linear CPTP errors
in the correlation space of
general resource states.
These results suggest that the application of
the theories of fault-tolerant quantum circuits
to the correlation space is not so straightforward.

TM and KF acknowledge supports by ANR (StatQuant, JC07 07205763)
and MEXT
(Grant-in-Aid for Scientific Research
on Innovative Areas 20104003), respectively.



\begin{thebibliography}{00}
\bibitem{one-way}
R. Raussendorf and H. J. Briegel, Phys. Rev. Lett. {\bf86}, 5188 (2001).
\bibitem{one-way2}
R. Raussendorf, D. E. Browne, and H. J. Briegel, Phys. Rev. A {\bf68},
022312 (2003).
\bibitem{one-way3}
R. Raussendorf, Ph.D. thesis, Ludwig-Maximillians Universit\"at
M\"unchen, 2003.

\bibitem{Gross1}
D. Gross and J. Eisert, Phys. Rev. Lett. {\bf98}, 220503 (2007).
\bibitem{Gross2}
D. Gross, J. Eisert, N. Schuch, and D. Perez-Garcia, Phys. Rev. A {\bf76},
052315 (2007).
\bibitem{Gross3}
D. Gross and J. Eisert, Phys. Rev. A {\bf82}, 040303(R) (2010).

\bibitem{Fannes}
M. Fannes, B. Nachtergaele, and R. F. Werner, J. Phys. A {\bf24}, L185 (1991).
\bibitem{Verstraete}
F. Verstraete, J. I. Cirac, and V. Murg, Adv. Phys. {\bf57}, 143 (2008).

\bibitem{Cirac}
J. I. Cirac and F. Verstraete, J. Phys. A: Math. Theor. {\bf42}, 504004 (2009).

\bibitem{Download}
J. M. Cai, W. D\"ur, M. Van den Nest, A. Miyake, and H. J. Briegel,
Phys. Rev. Lett. {\bf103}, 050503 (2009).

\bibitem{FM}
K. Fujii and T. Morimae, arXiv:1106.3377

\bibitem{Shor}
P. W. Shor, 
{\it Proc. of the 37th Symposium on Foundations
of Computing}, p.56 (IEEE Computer Society Press, 1996).


\bibitem{Aharonov}
D. Aharonov and M. Ben-Or,
{\it Proc. of the 29th Annual ACM Symposium on Theory of Computing},
p. 176 
(ACM Press, New York, 1998).

\bibitem{Kitaev}
A. Yu. Kitaev, Russian Math. Surveys {\bf52}, 1191 (1997).






\bibitem{Knill}
E. Knill, R. Laflamme, and W. H. Zurek, Proc. Roy. Soc. London, Ser A {\bf454},
365 (1998).

\bibitem{Terhal}
B. M. Terhal and G. Burkard,
Phys. Rev. A {\bf71}, 012336 (2005).


\bibitem{ND05}
M. A. Nielsen and C. M. Dawson,
Phys. Rev. A {\bf 71}, 042323 (2005).

\bibitem{AL06}
P. Aliferis and D. W. Leung,
Phys. Rev. A {\bf 73}, 032308 (2006).



\bibitem{Silva}
M. Silva, V. Danos, E. Kashefi, and H. Ollivier,
New. J. Phys. {\bf9}, 192 (2007).

\bibitem{Fujii1}
K. Fujii and K. Yamamoto,
Phys. Rev. A {\bf82}, 060301(R) (2010).
\bibitem{Fujii2}
K. Fujii and K. Yamamoto,
Phys. Rev. A {\bf81}, 042324 (2010).
\bibitem{recently}
Recently, a FT measurement-based quantum computation was considered
in Ref.~\cite{Li} for spin-$3/2$ and spin-$2$ resource states
which can be converted into the cluster state where
FT methods are available~\cite{topological1,topological2}.
\bibitem{Li}
Y. Li, D. E. Browne, L. C. Kwek, R. Raussendorf, and
T. C. Wei,
Phys. Rev. Lett. {\bf107}, 060501 (2011).


\bibitem{topological1}
R. Raussendorf and J. Harrington,
Phys. Rev. Lett. {\bf98}, 190504 (2007).
\bibitem{topological2}
R. Raussendorf, J. Harrington, and K. Goyal,
New J. Phys. {\bf9}, 199 (2007).

\bibitem{no-go}
J. Chen, X. Chen, R. Duan, Z. Ji, and B. Zeng,
Phys. Rev. A {\bf83}, 050301(R) (2011).



\bibitem{tricluster}
X. Chen, B. Zeng, Z. C. Gu, B. Yoshida, and I. L. Chuang,
Phys. Rev. Lett. {\bf102}, 220501 (2009).


\bibitem{Brennen}
G. K. Brennen and A. Miyake, Phys. Rev. Lett. {\bf101}, 010502 (2008).

\bibitem{AKLT}
I. Affleck, T. Kennedy, E. H. Lieb, and H. Tasaki, Comm. Math. Phys. {\bf115}, 477 (1988).

\bibitem{Caimagnet}
J. M. Cai, A. Miyake, W. D\"ur, and H. J. Briegel,
Phys. Rev. A {\bf82}, 052309 (2010).


\bibitem{Miyake2dAKLT}
A. Miyake, 
Ann. Phys. {\bf326}, 1656 (2011).



\bibitem{exclude}
We do not consider the case where
$A[1]\propto A[2]$,
since in this case we can reduce the dimension $d$ to $d-1$
by redefining the basis of the two-dimensional subspace
spanned by $|1\rangle$
and $|2\rangle$.


\bibitem{intuition}
This result is reasonable
since we physically do
a non-linear (or non-TP) operation.
If we consider this fact, it is surprising
that 
linear CPTP operations are implemented by 
doing non-linear (or non-TP) physical operations
on several resource states such as the cluster state
when physical operations are perfect!

\bibitem{details}
For details, see T. Morimae and K. Fujii,
arXiv: 1110.4182.

\bibitem{GrossH}
For
the AKLT-like resource state, $A[0]=X$,
$A[1]=XZ$, and $A[2]=H$,
which was proposed in Ref.~\cite{Gross1,Gross2},
it seems to be impossible to consider such a mixing strategy
due to the existence of the Hadamard $H$.








\end{thebibliography}
\end{document}